\documentclass[conference,10pt,table,a4paper]{IEEEtran}

\usepackage[utf8]{inputenc}
\usepackage{array} 
\usepackage{epsfig}
\usepackage{epstopdf}
\usepackage{times}
\usepackage{framed}
\usepackage{enumitem}
\usepackage{float}
\usepackage{etoolbox}
\usepackage{algpseudocode}
\usepackage{amssymb}
\usepackage{amsmath}
\usepackage{multirow}
\usepackage{url}
\usepackage{lipsum}
\usepackage{soul}
\usepackage{caption}
\usepackage{mathtools}
\usepackage{bbm}
\usepackage{booktabs}
\usepackage{xcolor}
\usepackage{makecell}
\usepackage{pifont}
\usepackage{subfigure}
\usepackage[export]{adjustbox} 
\usepackage{latexsym}
\usepackage{cite}
\usepackage[acronym]{glossaries}
\usepackage{multicol}
\usepackage{eurosym}
\usepackage{xspace}
\usepackage{pdfpages}
\usepackage{verbatim}
\usepackage{stfloats}
\usepackage{mathtools}
\usepackage[norelsize,boxruled]{algorithm2e}
\usepackage[normalem]{ulem}
\usepackage{svg}

\usepackage{blindtext}

\newcommand{\vv}[1]{\mathbf{#1}}

\newcommand{\Compl}{\mathbb{C}}
\newcommand{\vvh}[1]{\mathbf{#1}^\mathrm{H}}

\newcommand{\name}{D-RISA}

\newacronym{ula}{ULA}{uniform linear array}
\newacronym[plural=BSs, firstplural=base stations (BSs)]{bs}{BS}{base station}
\newacronym{aod}{AoD}{angle of departure}
\newacronym{ue}{UE}{user equipment}
\newacronym{los}{LoS}{line-of-sight}
\newacronym{nlos}{NLoS}{non-line-of-sight}
\newacronym{pla}{PLA}{planar linear array}
\newacronym{drl}{DRL}{deep reinforcement learning}
\newacronym[plural=RISs, firstplural=reconfigurable intelligent surfaces (RISs)]{ris}{RIS}{reconfigurable intelligent surface}
\newacronym{snr}{SNR}{signal-to-noise ratio}
\newacronym{mmwave}{mm-Wave}{millimeter-wave}
\newacronym{csi}{CSI}{channel state information}
\newacronym{sinr}{SINR}{signal-to-interference-plus-noise ratio}
\newacronym{sbr}{SBR}{shooting and bouncing rays}
\newacronym{ura}{URA}{uniform rectangular array}
\newacronym{uca}{UCA}{uniform circular array}
\newacronym{sncf}{SNCF}{Société nationale des chemins de fer français}
\newacronym[plural=CSs, firstplural=candidate sites (CSs)]{cs}{CS}{candidate site}
\newacronym[plural=TPs, firstplural=test points (TPs)]{tp}{TP}{test point}
\newacronym{dql}{DQL}{Deep Q-Learning}
\newacronym{dqfd}{DQfD}{Deep Q-learning from Demonstrations}
\newacronym{mrt}{MRT}{maximum ratio transmission}
\newacronym{fp}{FP}{Fractional Programming}
\newacronym{tdma}{TDMA}{time-division multiple access}
\newacronym{em}{EM}{electromagnetic}
\newacronym{soa}{SOA}{state-of-the-art}
\newacronym{iot}{IoT}{Internet of Things}
\newacronym{5g}{5G}{fifth-generation}
\newacronym{b5g}{B5G}{beyond-fifth-generation networks}
\newacronym{capex}{CAPEX}{capital expenditure}
\newacronym{ai}{AI}{Artificial intelligence}
\newacronym{es}{ES}{exhaustive search}
\newacronym{mimo}{MIMO}{multiple-input and multiple-output}
\newacronym{mlp}{MLP}{multilayered perceptron}
\newacronym{relu}{ReLU}{rectified linear unit}
\newacronym[plural=UAVs, firstplural=unmanned aerial vehicles (UAVs)]{uav}{UAV}{unmanned aerial vehicle}
\newacronym{itu}{ITU}{International Telecommunication Union}

\IEEEoverridecommandlockouts

\begin{document}

\setlength{\textfloatsep}{3pt}

\title{Unlocking Metasurface Practicality for B5G Networks: AI-assisted RIS Planning\\ {\large As accepted in the IEEE Global Communications Conference 2023}}

\author{
    \IEEEauthorblockN{Guillermo Encinas-Lago\IEEEauthorrefmark{1}\IEEEauthorrefmark{2}, Antonio Albanese\IEEEauthorrefmark{3}\IEEEauthorrefmark{4},
    Vincenzo Sciancalepore\IEEEauthorrefmark{1}, \\ Marco Di Renzo\IEEEauthorrefmark{2},
    Xavier Costa-P\'erez\IEEEauthorrefmark{1}\IEEEauthorrefmark{5}}
    \IEEEauthorblockA{
	\IEEEauthorrefmark{1}NEC Laboratories Europe, 69115 Heidelberg, Germany\\
	\IEEEauthorrefmark{2}Universit\'e Paris-Saclay, CNRS, CentraleSup\'elec, Laboratoire des Signaux et Syst\`emes, 91190 Gif-sur-Yvette, France\\
 	\IEEEauthorrefmark{3}Departamento de Ingeniería Telemática, University Carlos III of Madrid, 28911 Legan\'es, Spain\\
 	\IEEEauthorrefmark{4}Flyhound Co., 10019 New York, US \\
 	\IEEEauthorrefmark{5}i2cat Foundation and ICREA, 08034 Barcelona, Spain\\}
}

\maketitle

\IEEEpubid{\begin{minipage}{\textwidth}\ \\[12pt]
\copyright2023 IEEE.  Personal use of this material is permitted.  Permission from IEEE must be obtained for all other uses, in any current or future media, including reprinting/republishing this material for advertising or promotional purposes, creating new collective works, for resale or redistribution to servers or lists, or reuse of any copyrighted component of this work in other works.
\end{minipage}}

\IEEEpubidadjcol

\begin{abstract}

The advent of \glspl{ris} brings along significant improvements for wireless technology on the verge of \gls{b5g}.
The proven flexibility in influencing the propagation environment opens up the possibility of programmatically altering the wireless channel to the advantage of network designers, enabling the exploitation of higher-frequency bands for superior throughput overcoming the challenging \gls{em} propagation properties at these frequency bands.

However, \glspl{ris} are not magic bullets. Their employment comes with significant complexity, requiring ad-hoc deployments and management operations to come to fruition. In this paper, we tackle the open problem of bringing \glspl{ris} to the field, focusing on areas with little or no coverage. In fact, we present a first-of-its-kind \gls{drl} solution, dubbed as \name{}, which trains a \gls{drl} agent and, in turn, obtains an optimal \gls{ris} deployment. We validate our framework in the indoor scenario of the Rennes railway station in France, assessing the performance of our algorithm against \gls{soa} approaches. Our benchmarks showcase better coverage, i.e., $10$-dB increase in minimum \gls{snr}, at lower computational time (up to $-25$\%) while improving scalability towards denser network deployments. 

\end{abstract}

\IEEEpubidadjcol
\section{Introduction}
\label{s:introduction}
\glsresetall

Current social and industrial trends call for superior all-around network capabilities. Phenomena like the proliferation of smart devices or large-scale \gls{iot} deployments require wider bandwidths, coverage ubiquity, improved localization accuracy, and network deployments able to serve much more densely populated areas. This impelling need is directly shaping future technology standards, pushing the boundaries of network operators, which continuously investigate disruptive solutions to enable never-decreasing revenue streams while satisfying the customers' growing demand for network performance~\cite{Saad2020}.

Although the adoption of \gls{mmwave} frequencies is a key enabler of such use cases for \gls{5g} and \gls{b5g} networks, those frequencies present higher attenuation and poor propagation properties, requiring direct \gls{los} paths between \glspl{bs} and \glspl{ue} to compensate for such relevant power losses. \Glspl{ris} pave the way to substantial beamforming gains packing a massive number of configurable passive reflectors (e.g., built using varactor diodes) equipped with low-cost and low-complexity electronics~\cite{BOL20_ComMag}.  

The key value of \glspl{ris}---adaptive reconfigurability---allows focusing the impinging signals onto some desired directions and dynamically altering the reflection at will, while drawing minimal power, installation and maintenance costs compared to deploying active \glspl{bs}. This makes \glspl{ris} an effective candidate technology to tackle the mobile dead-zone problem in challenged scenarios, e.g., in classical indoor environments, at low \gls{capex}, by \emph{simply} steering \gls{em} waves impinging on \glspl{ris} towards sectors with little to no coverage, dramatically improving the \gls{sinr} experienced by \glspl{ue} in those areas~\cite{LiuY2021}.  

\IEEEpubidadjcol

However, \gls{ris} deployments still exhibit some practicality issues. As \glspl{ris} properly funnel the \gls{em} energy impinging on their surface carrying useful signal, they might also divert unwanted interference~\cite{Mursia_lwc2023}, reducing the \gls{sinr} gains when such effect is not accounted for. Furthermore, current \gls{ris} technology---with few notable exceptions employing self-configuring \glspl{ris}~\cite{albanese22}---hinges on an ad-hoc control channel to distribute the optimal reflection configurations to multiple \glspl{ris} managed by a common centralized controller. 

It is therefore crucial to contain the extension of such channel, which would increase complexity and deployment costs, thus limiting the agility of \gls{ris}-enabled networks. In this regard, advanced optimization techniques need to be developed and employed at the network planning stage, aiming at identifying the optimal trade-off among network coverage, interference reduction, and \glspl{ris} density~\cite{Ye2022,Ma2021}. In order to retain the low \gls{capex} benefits of these devices, the newly developed methods (as our deployment solution) for \glspl{ris} need to be also computationally inexpensive.

\subsection{Contributions} 
In this paper, we propose \name{}, namely Deep \gls{ris}-Aware network deployment and planning, as a novel \gls{drl}-based solution to enable practical installations of \glspl{ris} in the field by raising a \emph{digital twin} of the environment performed via a ray tracing simulation and a 3D model of the target area. 

We introduce a paradigm shift in the use of \gls{drl} to solve \gls{ris} deployment problems by not aiming at producing a \gls{drl} agent able to solve any problem instance, but rather tailoring the training of \name{} to the specific problem at hand, 
making it less demanding with respect to the available literature while avoiding the discretization of the solution space and thus increasing the solution scalability. 

In particular, the contributions made by this paper can be summarized as follows: we $i$) translate the \gls{ris} deployment problem into a \gls{drl} problem by means of the bespoke space and action spaces design, $ii$) identify the minimum \gls{snr} as a feasible metric to be computed by a custom-made ray tracing simulator to feed the \gls{drl} training process, $iii$) make use of the \name{} \gls{drl} agent training phase as an exploration tool to find the best deployment solution, avoiding the need for a complete viable agent, $iv$) showcase the rapid convergence of the agent solution and $v$) benchmark \name{} against the \gls{soa} as well as exhaustive approaches, demonstrating outstanding performance while emulating the indoor scenario of a real environment, namely the Rennes railway station in France. 

\begin{table}[h!]
\caption{{List of abbreviations}}
\label{tab:acronyms}
\centering
\resizebox{\linewidth}{!}{
\begin{tabular}{cc|cc}
\hline
5G	&	fifth-generation	&	B5G	&	beyond-fifth-generation networks	\\
AI	&	artificial intelligence	&	CAPEX	&	capital expenditure	\\
AoD	&	angle of departure	&	CSI	&	channel state information	\\
BS	&	base station	&	DQfD	&	deep Q-learning from demonstrations	\\
CS	&	candidate site	&	DRL	&	deep reinforcement learning	\\
DQL	&	deep Q-learning	&	ITU	&	International Telecommunication Union	\\
EM	&	electromagnetic	&	MIMO	&	multiple-input and multiple-output	\\
ES	&	exhaustive search	&	MLP	&	multilayered perceptron	\\
FP	&	fractional programming	&	MRT	&	maximum ratio transmission	\\
IoT	&	internet of things	&	ReLU	&	rectified linear unit	\\
LoS	&	line-of-sight	&	RIS	&	reconfigurable intelligent surface	\\
mm-Wave	&	millimeter-wave	&	SBR	&	shooting and bouncing rays	\\
NLoS	&	non-line-of-sight	&	SINR	&	signal-to-interference-plus-noise ratio	\\
PLA	&	planar linear array	&	SNCF	&	Société nationale des chemins de fer français	\\
SNR	&	signal-to-noise ratio	&	TDMA	&	time-division multiple access	\\
SOA	&	state-of-the-art	&	UAV	&	unmanned aerial vehicle	\\
TP	&	test point	&	UCA	&	uniform circular array	\\
UE	&	user equipment	&	URA	&	uniform rectangular array	\\
ULA	&	uniform linear array	&		&		\\

\hline
\end{tabular}
}
\end{table}

\section{Framework overview}

\label{s:overview}
\begin{figure}[t]
        \center
        \includegraphics[width=\linewidth]{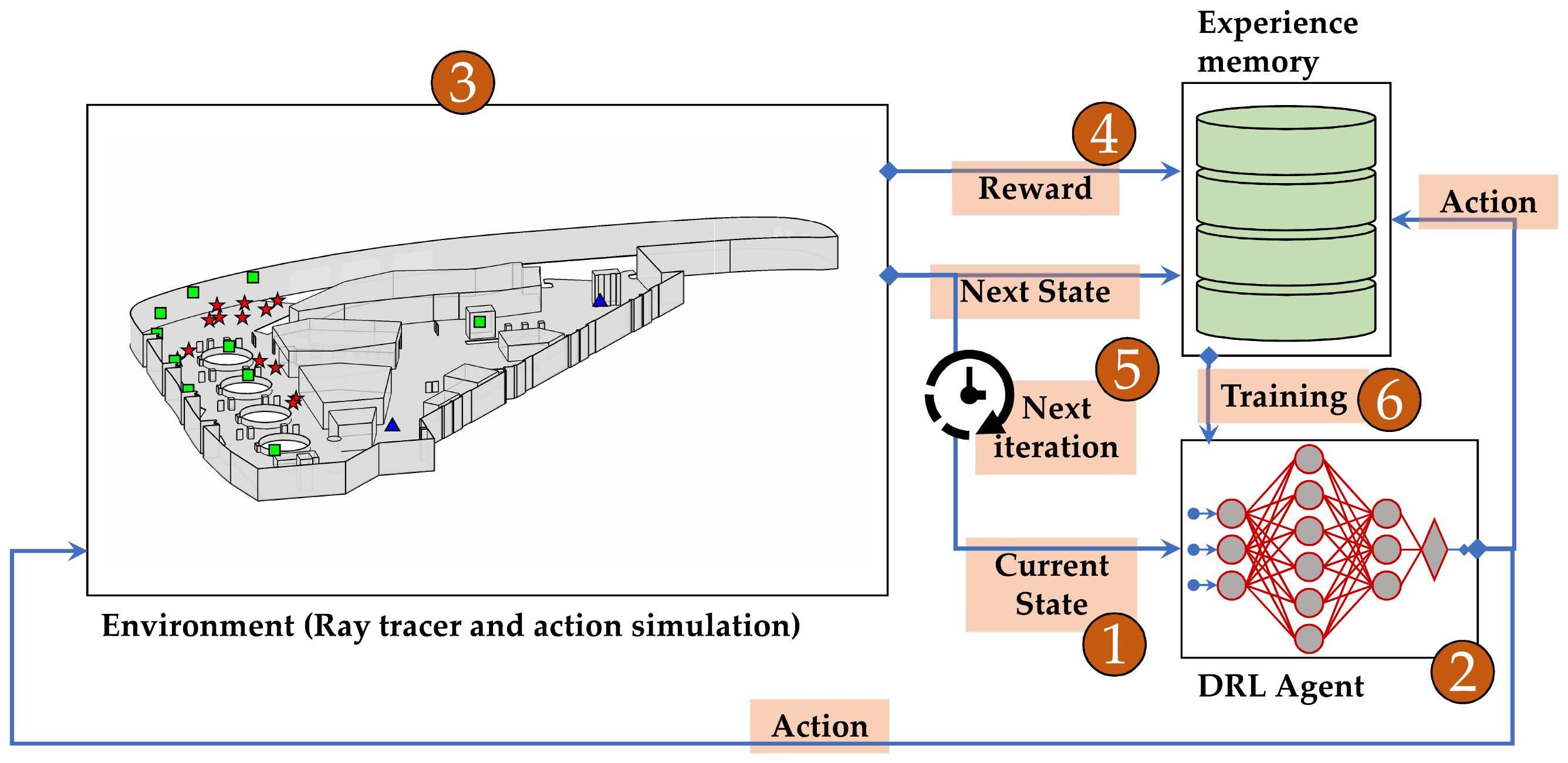}
        \caption{\name{}'s building blocks showing the training process for a given frame. In the 3D map, the \glspl{cs} for the \gls{ris} deployment are shown in green, the \glspl{bs} in blue, and \glspl{tp} in red.}
        \label{fig:solution_schematic} 
\end{figure}

Hereafter, we tackle the problem of efficiently deploying \glspl{ris} in a known environment. We extend a currently existing network deployment with \glspl{ris} to eliminate coverage dead zones. We consider the environment of the Rennes railway station hall, operated by the French \gls{sncf}\footnote{We build the 3D model from the architectural data, and the information provided by the network operator servicing the area.}, and visualized in Fig.~\ref{fig:solution_schematic}. 
In the present work, we try to alleviate the dead zone problem. The current deployment of \glspl{bs} in the Rennes station has several areas with connectivity problems due to the low \gls{snr} achieved. We identify and set \glspl{tp} on those locations and alleviate the problem with our proposed solution.
Specifically, we make use of a custom-made ray tracing engine to simulate any proposed deployment in the scenario while accounting for its current network infrastructure, provided by a major European network operator. As the ray tracer outputs the values of the objective function (the minimum \gls{snr} across all the \glspl{tp} defined, as described in Section~\ref{s:ris_planning}) we train a \gls{drl} solution by using them as rewards.

In Fig.~\ref{fig:solution_schematic}, we showcase the main building blocks of our proposed solution, itemizing the operations required for a single training cycle iteration, which is later detailed in Section~\ref{s:solution}. Here we only assess the main components of \name{}, namely $i$) a 3D model of the scenario, $ii$) a ray tracer running on such a scenario and capable of evaluating candidate deployment solutions, $iii$) a model for the \gls{ris} gain, which we use in combination with the ray tracer to numerically evaluate any solution, $iv$) a \gls{dql} agent, which takes as input a given solution and decides the best among the different potential changes on it, with the long-term goal of improving the considered objective, and $v$) the training process of the mentioned agent, according to the principles of \gls{dqfd}~\cite{hester2018deep}. We would like to underline that \gls{dql} is only one flavor of \gls{drl} and that \name{} can be readily extended to other \gls{drl} techniques with minimal modifications.

{\bf 3D model.} The first component of the digital twin---the 3D model of our target scenario---is based on the architectural plans of the Rennes railway station, which is currently equipped with 2 deployed \glspl{bs}. Similarly to any deployment situation, network operators must comply with regulatory and physical restrictions, thus we consider a predefined number of \glspl{cs} at which \glspl{ris} might be deployed\footnote{We would like to underline that there is no intrinsic limitation on the \glspl{cs} set, so in principle they could cover the entire target area.}. In this work, we handpick the set of \glspl{cs}, making sure that the installation of \glspl{ris} at those locations is materially feasible, and ensuring \gls{los} between \glspl{cs} and both existing \glspl{bs} and \glspl{tp}. For the selection of \glspl{tp}, we focus on the areas experiencing the worst coverage problems under the current \glspl{bs}-only network deployment. These \glspl{tp} serve as probe points for the objective and represent \glspl{ue} to be served after deployment, matching their expected distribution\footnote{Please note that \name{} does not make any assumption on the specific \glspl{ue} or \glspl{tp} distributions.}~\cite{albanese2022loko}.

{\bf Ray tracer.} The second component of the digital twin---the ray tracing simulator---relies on a commercial \gls{sbr} engine to evaluate any given deployment within a 3D model~\cite{Brem2015}. To this aim, it computes the possible paths, both \gls{los} and \gls{nlos}, between any \gls{ris} and any \gls{tp} or \gls{bs}. Then, it derives their geometric information (i.e., the \gls{aod} from the \gls{ris}), which is fundamental for assessing the end-to-end path losses and so the objective function for the given deployment. 
Thanks to the use of the ray tracer, our approach does not need to make assumptions on the \gls{bs}-\gls{ris}-\gls{tp} channels, as they are individually simulated: any fading is therefore taken into account in the solution.  

{\bf \gls{ris} gain model.} As mentioned in Section~\ref{s:introduction}, we face a chicken-egg problem. To decide the best deployment of \glspl{ris} we need to start from a known configuration, whereas the optimal configurations can be computed only upon fixing the deployment and the \gls{bs}-\gls{ris}-\gls{tp} associations. Here we anticipate that we overcome the problem by considering a cellular-like architecture in which each \gls{ris} serves one contiguous subarea, thus reducing the scope of inter-\gls{ris} interference to the edges of said subareas. Then, to achieve full decoupling of the \glspl{bs} and \glspl{ris} beamforming optimization from the planning problem, we assume that each \gls{bs} can be associated to multiple \glspl{ris}, while operating in a \gls{tdma} fashion, namely radiating towards one of them at a time, thereby allowing \gls{mrt} precoding at the \glspl{bs}. In these conditions, the \gls{ris} gain model can be obtained by means of beam broadening and flattening as per~\cite{Lu2021}. {It is worth noting that the approach detailed in~\cite{Lu2021} assumes continuous phase shifts for the RISs, and hence we inherit that assumption. Nevertheless, this technique, based on the sub-division of the device in sub-\glspl{ris} that act in coordination, can be applied to discrete phase models for the \glspl{ris}, with minimal performance losses due to the phase discretization~\cite{di2020hybrid}. Additionally, we design \name{} to be completely decoupled from the \gls{ris} model used to compute the \gls{snr}, and hence, the gain of the \gls{ris}.} We further analyze the \gls{ris} planning problem in Section~\ref{s:ris_planning}.    

{\bf \gls{dql} agent.} In Section~\ref{s:solution}, we design the \gls{dql} agent used to explore the space of possible deployments. Our \gls{dql} agent considers deployments and identifies the best actions to improve them (e.g., placing an \gls{ris} at a specific \gls{cs}). Such improvement is not obtained by means of a mere greedy approach. On the contrary, the \name{} agent attempts to evolve the solution to the one maximizing the objective. By training the agent, we explore possible deployments combining acquired knowledge and random exploration, and find the best deployment solution upon stopping the training of the agent when the related criteria are met.

\section{RIS planning}

\label{s:ris_planning}

\begin{figure}[t]
        \center
        \includegraphics[width=\linewidth]{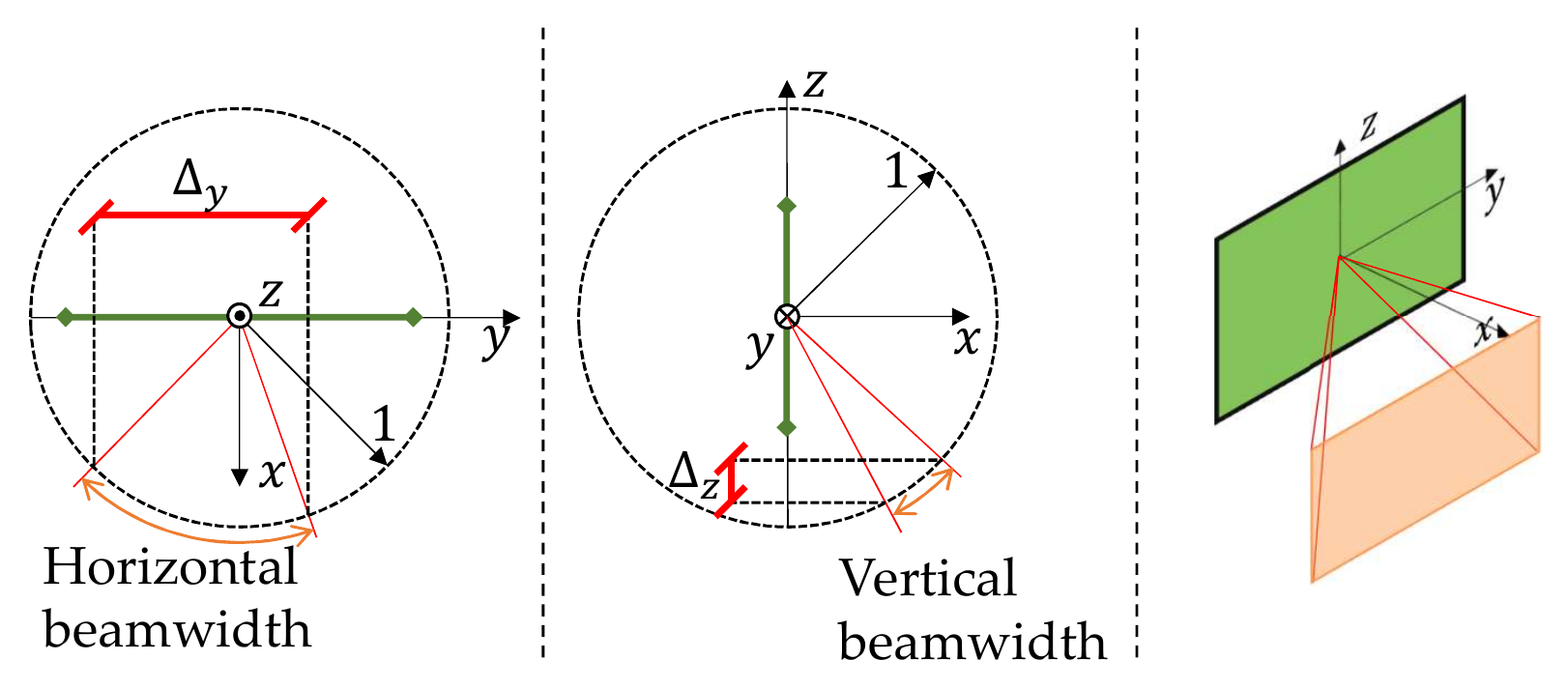}
        \caption{Geometry of the \gls{ris} gain model.}
        \label{fig:model_geometry} 
\end{figure}

We consider an \gls{ris}-enabled network comprising of $L$ \glspl{ris} deployed to aid a pre-existing deployment of $M$ \glspl{bs} in order to reduce the areas with limited or no coverage.
Each \gls{bs} is a \gls{ula} of $N_b$ antennas, and each \gls{ris} is a \gls{pla} of $N_r = N_{h} \times N_v$ reflective elements, where $N_h$ and $N_v$ denote the number of elements in the horizontal and the vertical directions, respectively. A symbol received at the \gls{ue} through the cascaded channel \gls{bs}-\gls{ris}-\gls{ue} is formulated according to the channel and the \gls{ris} element configuration, as the following:
\begin{equation}\label{eq:y}
    y = \left(\vvh{h} \vv{\Theta} \vv{G}\right)  \,  \vv{w} s + n \in \Compl,
\end{equation}
where $\vvh{h}$ represents the Hermitian transposition of the channel between the \gls{ris} and the receiver,  $\vv{\Theta} = \mathrm{diag}\left(\alpha_{1} e^{j\phi_{1}}, \dots, \alpha_{N} e^{j\phi_{N}}\right)$ with $\phi_{i}$ and $\alpha_{i}$ representing the phase shift and magnitude of the $i$-th element of the \gls{ris}, $\vv{G}$ is the channel between the elements of the \gls{ris} and the individual antennas of the \gls{bs}, $\vv{w}$ is the vector of transmit weights of the \gls{bs} antennas while $s$ is the transmitted symbol, and $n$ is the additive white Gaussian noise term distributed as $\mathcal{CN}(0,\sigma^2)$. {Since we do not specialize any mathematical expressions based on the distribution of the \gls{bs} elements, the assumption of \gls{ula} shape can be generalised to other arrangements, including \glspl{uca}, \glspl{ura}, as long as they can be adequately modeled and computed.} For our model, we rely on a ray tracer engine, overcoming the need for channel estimation and models.
As mentioned in Section~\ref{s:overview}, we configure the \glspl{ris} to serve their respective subarea using the beam broadening and flattening technique proposed in~\cite{Lu2021}. We subdivide each \gls{ris} in subsections so that they interfere constructively to obtain a beam of constant gain and controlled width. The gain of such beam $g$ depends only on the number of elements and shape of the \gls{ris}, namely 
\begin{equation}
    g(N_h, N_v, \Delta_{y}, \Delta_{z}) \propto \frac{N_h^2}{\Delta_{y}N_h \delta} \frac{N_v^2}{\Delta_{z}N_v \delta}, \label{eq:g1}
\end{equation}

where $\delta$ is the ratio between the \gls{ris} inter-element distance and the signal wavelength, and $\Delta_{y}$ and $\Delta_{z}$ are the spatial frequencies comprised in the beam in both directions, as illustrated in Fig.~\ref{fig:model_geometry}.
Using this gain and the antenna models of the \glspl{bs} and \glspl{ue}, we compute the \gls{snr} of the individual \glspl{tp} under a given deployment.

We can now formulate the objective function as follows 
\begin{align}
    \max_{\mathcal{S}} \min_{\vv{u}_t} \mathrm{SNR}(\vv{u}_t),
\end{align}
in which $\mathcal{S}$ denotes the solution space, defined as the set of possible \gls{bs}-\gls{ris}-\gls{tp} associations and possible \gls{ris} deployments, while $\vv{u}_t$ represents the coordinates of test point $t$, with $t=0,\dots,T-1$, where $T$ is the number of considered test points. The solution space $\mathcal{S}$ is thoroughly analyzed in Section~\ref{s:solution}. By indicating the number of \glspl{cs} for the \gls{ris} deployment as $N \geq L$, the considered problem consists of selecting $L$ out of $N$ \glspl{cs} to deploy the \glspl{ris} and identifying the associations with \glspl{bs} and \glspl{tp}.

A conventional approach to tackle the problem of optimally deploying \glspl{ris} is to extend known algorithms used for the deployment of non-\gls{mimo} and \gls{mimo} networks. The problem can then be written as an optimization problem and tackled with non-convex optimization tools. It can be relaxed via \gls{fp} to derive a tractable form, obtaining approximate solutions after binary rounding~\cite{albanese2021rennes}. However, such methods are model-based, and are not able to consider the actual propagation environment. On the other hand, \gls{drl}, being data-driven, can leverage large measurement sets to tailor the deployment to the \gls{em} properties of the target area. Ray tracing simulations can generate the training data at the planning stage using 3D models.

{\bf Q-learning.} \name{} relies on a  model-free \gls{drl} technique, \gls{dql}. It operates in a space of possible states $\mathcal{S}$, and a space of actions $\mathcal{A}$ that can be taken in each state. Each action taken evolves the state into a new one, and produces a reward. A quantity Q is associated with each pair state-action, based on immediate and potential future rewards. The policy of the algorithm is to choose, for each state, the action with the highest Q value.

To train a Q-learning agent, we feed it with states from space $\mathcal{S}$, and allow it to make random choices (\textit{exploration}) or informed choices based on its past learning (\textit{exploitation}). The atomic element of the training process is the frame. A frame considers an initial state $s \in \mathcal{S}$, an action $a \in \mathcal{A}$ chosen by the Q-learning agent or randomly picked, and the state evolution to the next state $s' \in \mathcal{S}$, producing a reward $r(s, s') \in \mathbb{R}$. The balance between exploration and exploitation varies with time, beginning with a preponderance of exploration to end mostly with exploitation. To randomize the individual decisions, if the total number of frames for a training run is predetermined to be $F$, at a given frame $f$, an exploratory action happens with probability $P_{explore} = 1-(f/F)$ and an exploitative action with a probability $P_{exploit} = f/F$.
After each training frame, Q-learning updates the associated value $Q(s,a)$ as

\begin{align}
 \displaystyle Q_{new}(s,a) =& 
 \,(1 - \alpha_q) \, Q_{old}(s,a)  \nonumber\\ 
 & + \alpha_q \left[ {r(s, s') + \gamma_q \, {\max_{a \in \mathcal{A}}Q(s', a)}} \right]  
, \label{eq:q_table_evolution}
\end{align}

where $\alpha_q$ determines the training speed, $\gamma_q$ determines how much potential future rewards affect the action selection, and $Q_{old}$ and $Q_{new}$ are the existing and updated values for $Q(s,a)$.

\section{Solution by space exploration}
\label{s:solution}

We describe here in detail the \name{} \gls{dql} agent. The structure of \gls{dql} builds upon the conventional Q-learning by substituting the Q-table with a neural network and adapting the training process. This way, the agent is not restricted to finite or discrete state and action spaces and avoids the large memory requirements of Q-learning. Additionally, the neural network learns from past data and infers patterns, even when encountering situations that have never been experienced before.

Analogously to Q-learning, we perform the training of the agent with a series of frames, grouped into episodes: an episode begins with a random initial state (no. 1 in Fig.~\ref{fig:solution_schematic}) and evolves as a consequence of the actions taken (no. 2 in Fig.~\ref{fig:solution_schematic}). For each action taken, the environment  produces a reward  (no. 3 and 5 in Fig.~\ref{fig:solution_schematic}). The episode ends when the \gls{dql} agent chooses the stop action or when the maximum number of frames is reached, as described in the following. \name{} follows the same exploration-exploitation strategy as Q-learning.

For each frame we record the initial state, the action taken, the final state and the reward in the experience memory buffer (no. 4 in Fig.~\ref{fig:solution_schematic}). With a predetermined frequency $T_T$, the neural network $\mathit{NN}$ that substitutes the Q-table is retrained using a random selection of those records (no. 6 in Fig.~\ref{fig:solution_schematic}). To compute the Q-values necessary to do so, we employ a copy $\mathit{NN}'$ of $\mathit{NN}$ to predict an estimation of the future rewards for each data instance of the selection of records from the buffer, namely 

\begin{equation}
 \displaystyle {\max_{a \in \mathcal{A}}Q(s', a)} \approx \max_{a \in \mathcal{A}}\mathit{NN}'(s',a).    \label{eq:future_rewards}
\end{equation}

Hence, we define the reference Q-values for each selected record as

\begin{equation}
 \displaystyle Q_{ref}(s,a) \triangleq r(s, s') + \gamma_q \, {\max_{a \in \mathcal{A}}\mathit{NN}'(s',a)}.   \label{eq:q_ref_values}
\end{equation}

We define a loss function to reward the ability of the neural network to mimic the $Q_{ref}$ values derived from the selected records. We compute the gradient of such loss function with respect to the trainable weights of the neural network, and we update the weights in the gradient direction. The neural network $\mathit{NN}'$ is updated by copying $\mathit{NN}$ with a frequency $T_{ST}^{-1}$, with $T_{ST} \gg T_T$. This way we prevent the instability of the training process, which makes use of the neural network recursively, since retraining the network may affect the estimation of future rewards and, in turn, affect the whole training process.
Using this method to navigate the state space, the \name{} agent is able to find viable deployment solutions. For allowing the operation of the agent, we formulate the solution space $\mathcal{S}$ and action space $\mathcal{A}$ as follows.

\textbf{Solutions space.} An individual state (or solution) $s \in \mathcal{S}$ consists of deployment and association. The deployment is formulated as a binary vector $\vv{x} = \{0,1\}^N$, where $x_n = 1$ represents a deployed \gls{ris} at \gls{cs} $n$. The association is translated into a binary matrix $Y \in \{0,1\}^{L \times T}$, where $y_{l,t}=1$ indicates that the deployed \gls{ris} $l$ provides coverage to the \gls{tp} $t$. Concatenating the binary deployment vector $\vv{x}$ with the individual rows of the association matrix $Y$, we obtain a binary representation of the state that the agent can handle. 

\begin{figure}[t!]
    \centering  
    \subfigure[\gls{snr} without \glspl{ris}]
    {
        \includegraphics[clip,width=.47\linewidth, trim = {0cm 10cm 0cm -0.5cm}]{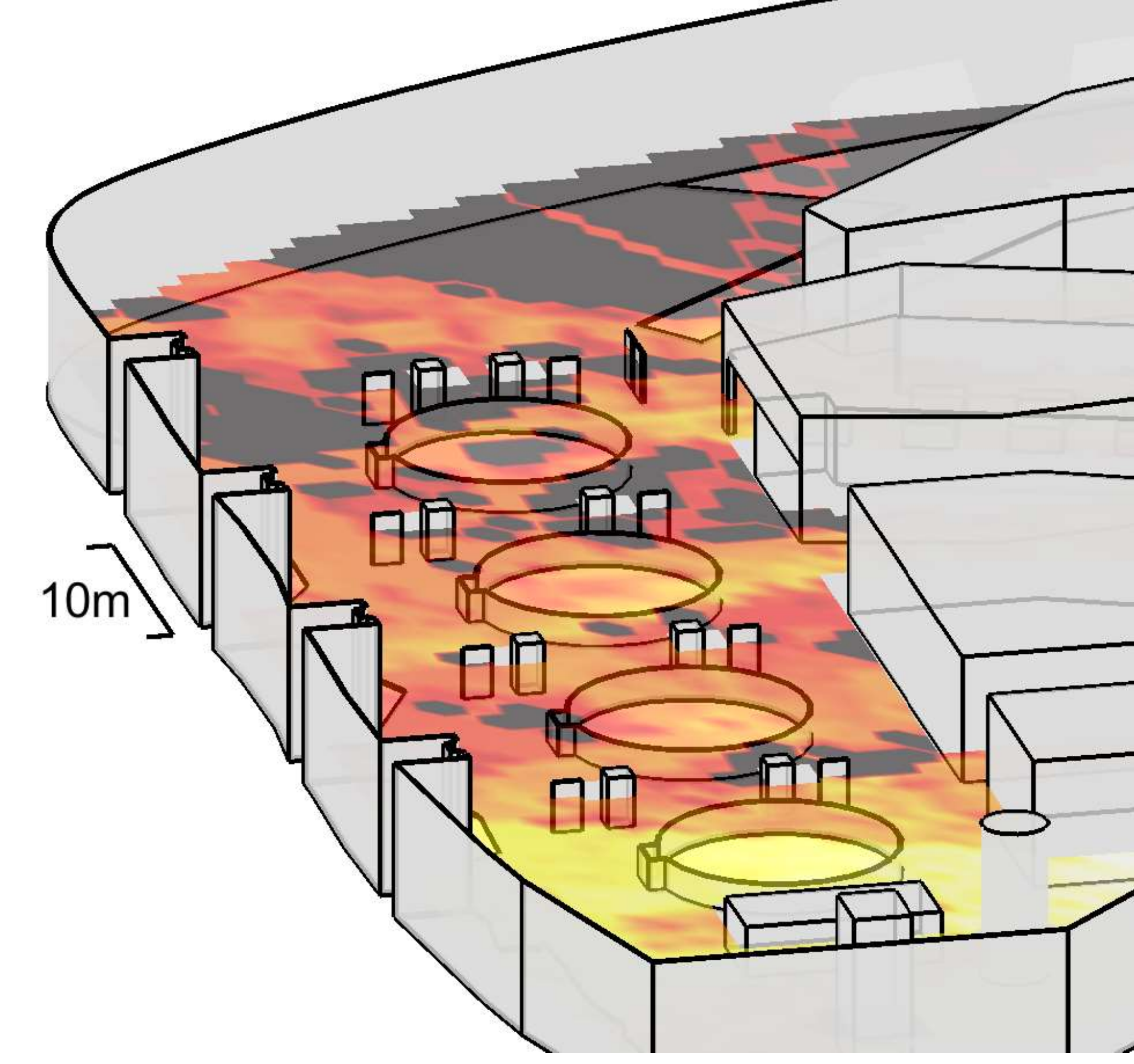}
        \label{fig:3D_heatmap_no_ris}
    } \hspace{-0.3cm}
    \subfigure[\gls{snr} with 6 \glspl{ris}]
    {
        \includegraphics[clip,width=0.47\linewidth, trim = {0cm 10cm 0cm 0cm}]{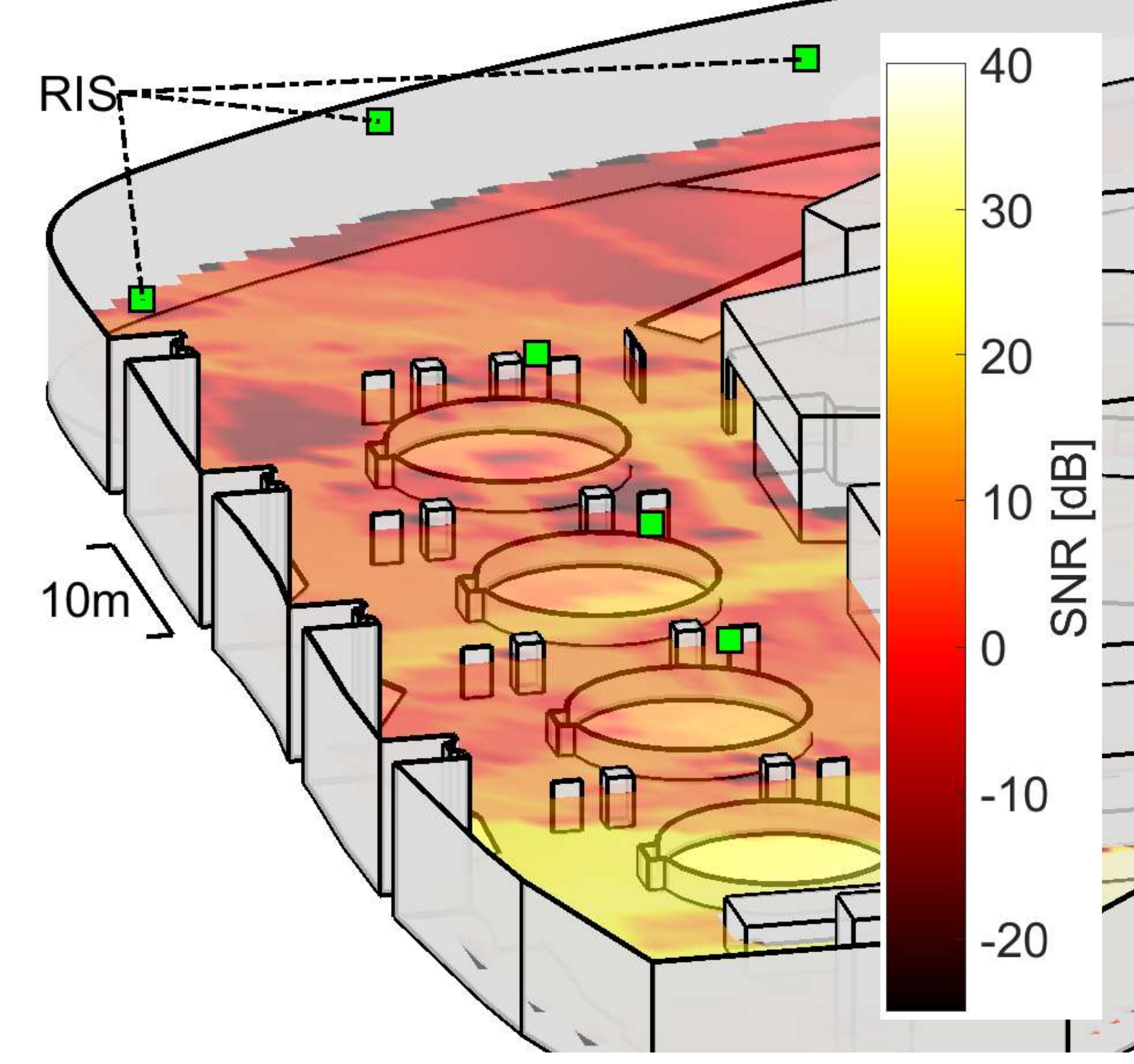}
        \label{fig:3D_heatmap_w_ris}
    }
    \caption{\gls{snr} analysis considering (a) the current arrangement and (b) the \gls{snr} achievable with $L = 6$ \glspl{ris}.
    }
    \label{fig:heatmap_comparison_rennes}
\end{figure}

\textbf{Actions space.} In \name{}, we define the following actions: $i)$ place a deployed \gls{ris} at a chosen \gls{cs}, $ii)$ remove a deployed \gls{ris} from a chosen \gls{cs}, $iii)$ change the association of a chosen \gls{tp} to the next available \gls{ris}, $iv)$ change the association of a chosen \gls{tp} to the previous available \gls{ris}, $v)$ end the current episode. 

{
\textbf{Alternative deployments.} Current \gls{soa} \gls{ris} prototypes, as the one presented in~\cite{rossanese2022designing}, are expected to have highly affordable mass-producing costs. Other techniques, such as smart skins, can attain similar properties with even simpler implementations, cheaper maintenance, and no electric consumption for operation, at the cost of no reconfigurability~\cite{flamini2022toward}. In this work, we assume very limited capabilities for the \gls{ris} to be able to reconfigure itself, either due to time or energy constraints. This flexibility allows \name{} to be applied to passive and static reflection technologies, such as smart skins, as well. The benefits of the capability of \glspl{ris} to be reconfigurable, in order to meet changing demands and propagation conditions, need to be evaluated in comparison to the traditional option of deploying additional \glspl{bs} or repeaters, which offers the best performance at the highest cost, or installing \glspl{ris} for improving the performance while retaining versatility, or even choosing the most affordable smart skin devices but accepting the lack of reconfigurability~\cite{flamini2022toward}.   
}

\section{Performance evaluation}
\label{s:performance}

\textbf{Simulation settings.} We consider the Rennes railway station scenario{. \name{} does not impose a restriction on the existing infrastructure, and increasing the number of pre-deployed \glspl{bs} has a negligible impact on the computational costs. Nevertheless, for this work, we reproduce in the ray tracer the currently existing installation of radio equipment in the location by the operator. As a result, we are able to ensure that the actual coverage problems are reproduced in the simulations, and we can obtain practically useful and applicable solutions. The considered scenario is} characterized by $M=2$ existing \glspl{bs} with a transmit power $P = 28$ dBm operating at $f=26$ GHz, and select $N = 10$ \glspl{cs}, and $T = 13$ \glspl{tp}. 
The coverage of the area, the positions of the existing \glspl{bs} and their properties are provided by the European major network operator serving the station.
The \glspl{tp} are scattered in the area, focusing on the sectors wherein existing \glspl{bs} are not able to provide adequate coverage, while the \glspl{cs} are handpicked in architecturally suitable places, both in areas directly surrounding the \glspl{tp} and outside the immediate vicinity, as depicted in Fig.~\ref{fig:solution_schematic}. The ray tracer uses a 3D model of the station, which reproduces the most prominent architectural features of the environment. The model is composed of a total of $579$ triangular surfaces. It uses the \gls{sbr} method to find all possible paths between any pair of given points. For simplicity, we use a single material for the whole environment as a reference to compute the losses at each reflection. The material properties are in agreement with the \gls{itu} values for the permittivity and conductivity of concrete at $f = 26$ GHz, i.e., real part of relative permittivity $Re(\varepsilon_r) = 5.31$, and conductivity $\sigma = 0.4557$ S/m~\cite{series2015effects}. To compute the overall effect of the deployment of \glspl{ris} in our scenario beyond the chosen test points, we also analyse the whole surface at a height of $1.5$ m. To produce such heatmaps, as depicted in Fig.~\ref{fig:heatmap_comparison_rennes} instead of the gain model given in~\eqref{eq:g1}, we model the \glspl{ris} as isotropic scatterers. We can see the difference between the current deployment without \glspl{ris} in Fig.~\ref{fig:3D_heatmap_no_ris} and with \glspl{ris} in Fig.~\ref{fig:3D_heatmap_w_ris}.

\begin{figure}[t]
        \center
        \includegraphics[width=1.0\linewidth, trim = {0cm 0cm 0cm -0.5cm}]{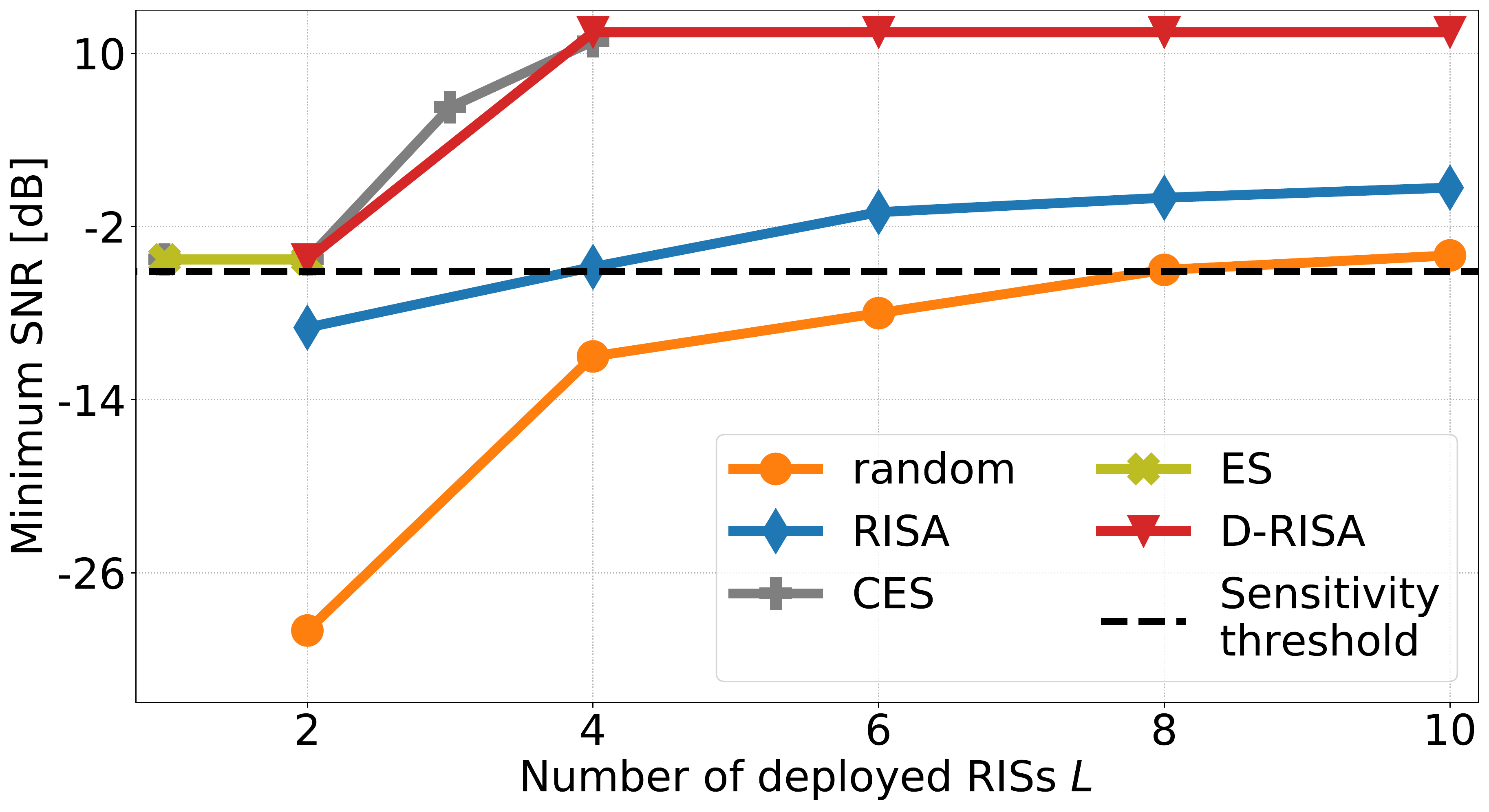}
        \caption{\name{} performance against benchmark methods and \gls{soa} RISA~\cite{albanese2021rennes} for a different number of deployed \glspl{ris} in terms of minimum \gls{snr}.}
        \label{fig:SNR} 
\end{figure}

\textbf{\gls{dql} hyperparameters.} 
We employ a \gls{mlp} neural network design with $2$ hidden, fully connected layers of $32$ neurons activated via a \gls{relu} function. The learning rate and the future reward reduction are respectively set to $\eta = 0.00025$ and $\gamma_q = 0.99$, which are optimized by means of the Adam algorithm~\cite{kingma2014adam}. The training period of the primary network $\mathit{NN}$ is $T_T = 4$ complete frames and the period between updates of the secondary network $\mathit{NN}'$ (i.e., by copying $\mathit{NN}$) is $T_{ST} = 500$ complete frames. The simulation settings and the \gls{dql} parameters are listed in Table~\ref{tab:parameters}.
\begin{table}[h!]
\caption{Simulation parameters}
\label{tab:parameters}
\centering
\resizebox{\linewidth}{!}{
\begin{tabular}{cc|cc|cc}
\textbf{Parameter} & \textbf{Value} & \textbf{Parameter} & \textbf{Value} & \textbf{Parameter} & \textbf{Value}\\  
\hline
\rowcolor[HTML]{EFEFEF}
$P$ & $28$ dBm & $f$ & $26$ GHz & $N$ & $10$ \\
$N_h$ & $347$ & $N_v$ & $175$ & $N_b$ & $2$ \\
\rowcolor[HTML]{EFEFEF}
$M$ & $2$  &  $Re(\varepsilon_r)$ & $5.31$ & $\sigma$ & $0.4557$ S/m\\
$T$ & $13$ & $\gamma_q $&$ 0.99$ &  $T_T $&$ 4$ \\
 \rowcolor[HTML]{EFEFEF}
$\eta$ & $0.00025$ & $T_T $&$ 4$ & $T_{ST} $&$ 500$  \\
\hline
\end{tabular}
}
\end{table}

\textbf{Benchmark.} We apply the proposed solution to the described scenario, and put it side-by-side with the following benchmarks: $i)$ the solution obtained by \gls{es}, $ii)$ the solution obtained by \gls{es} considering clusters of \glspl{tp} instead of individual \glspl{tp}, drastically reducing the number of possible associations and allowing us to extend the \gls{es} up to $L = 4$, $iii)$ the solution obtained by means of the \gls{soa} RISA~\cite{albanese2021rennes}, which is based on \gls{fp}, and $iv)$ a statistical average of random solutions as previously obtained in the same scenario by~\cite{albanese2021rennes}. As shown in Fig.~\ref{fig:SNR}, \name{} outperforms all other solutions and matches the \gls{es} in terms of minimum \gls{snr} at a much lower computational cost. For reference, we also depict the \gls{snr} threshold corresponding to the typical receiver sensitivity. 

\begin{figure}[t]
        \center
\includegraphics[width=1.0\linewidth, trim = {0cm 0cm 0cm 0cm}]{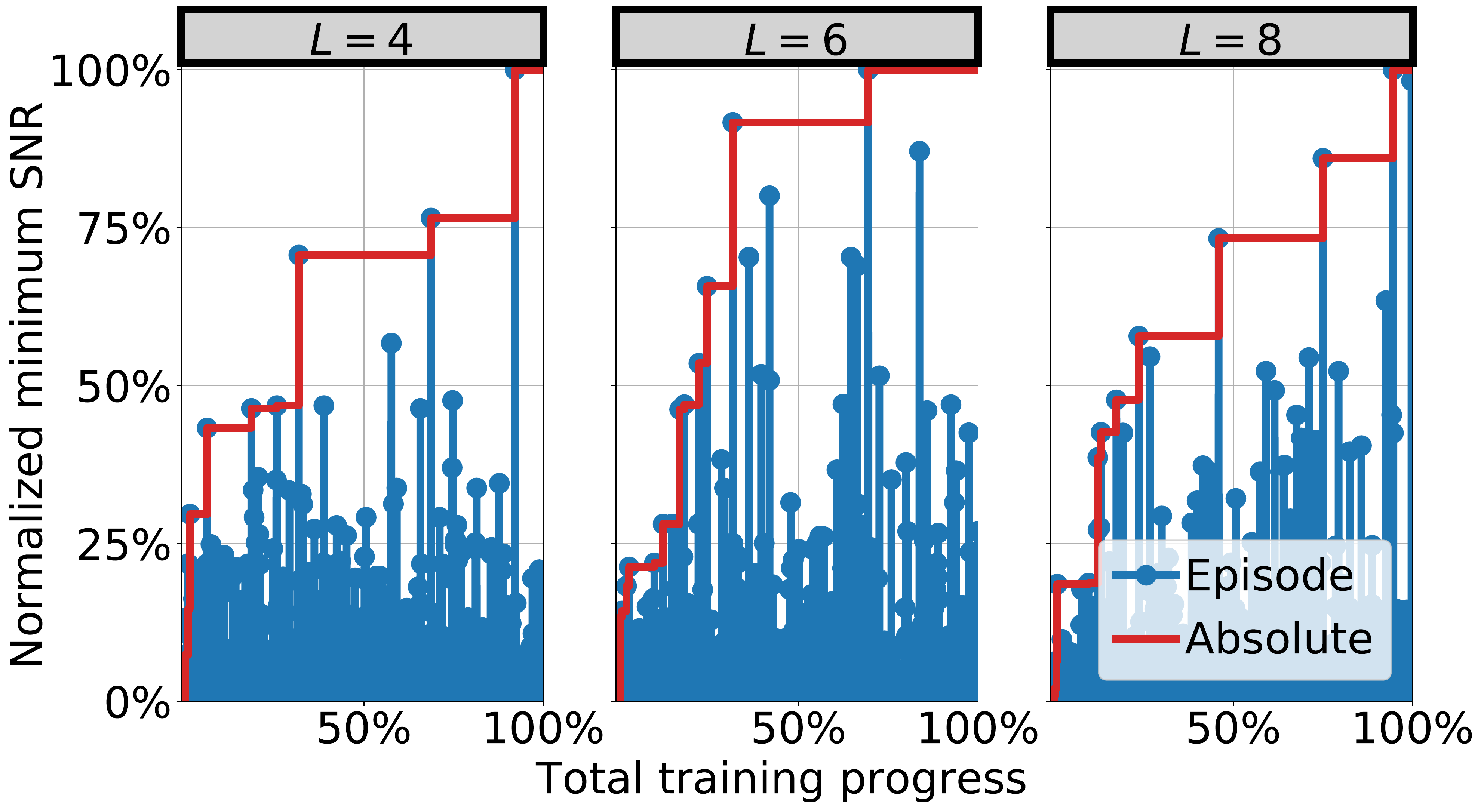}
        \caption{Solution evolution in terms of normalized minimum \gls{snr} for a different number of deployed \glspl{ris}.}
        \label{fig:best_in_episode} 
\end{figure}

\textbf{\name{} training.} Fig.~\ref{fig:best_in_episode} illustrates the \name{} agent training for $L \in \{4,6,8\}$ by showcasing the best normalized minimum \gls{snr} during each episode and the best solution found at each point in the training process. We observe an increasing trend as the agent keeps building its ability to improve the deployments. As the training process ends with an exploitative behavior (as per Section~\ref{s:solution}), the high values of the best normalized minimum \gls{snr} during each episode suggest that the agent has learned a successful strategy to solve the problem even without exploratory actions.

\begin{figure}[t]
        \center
\includegraphics[width=1.0\linewidth, trim = {0cm 0cm 0cm 0cm}]{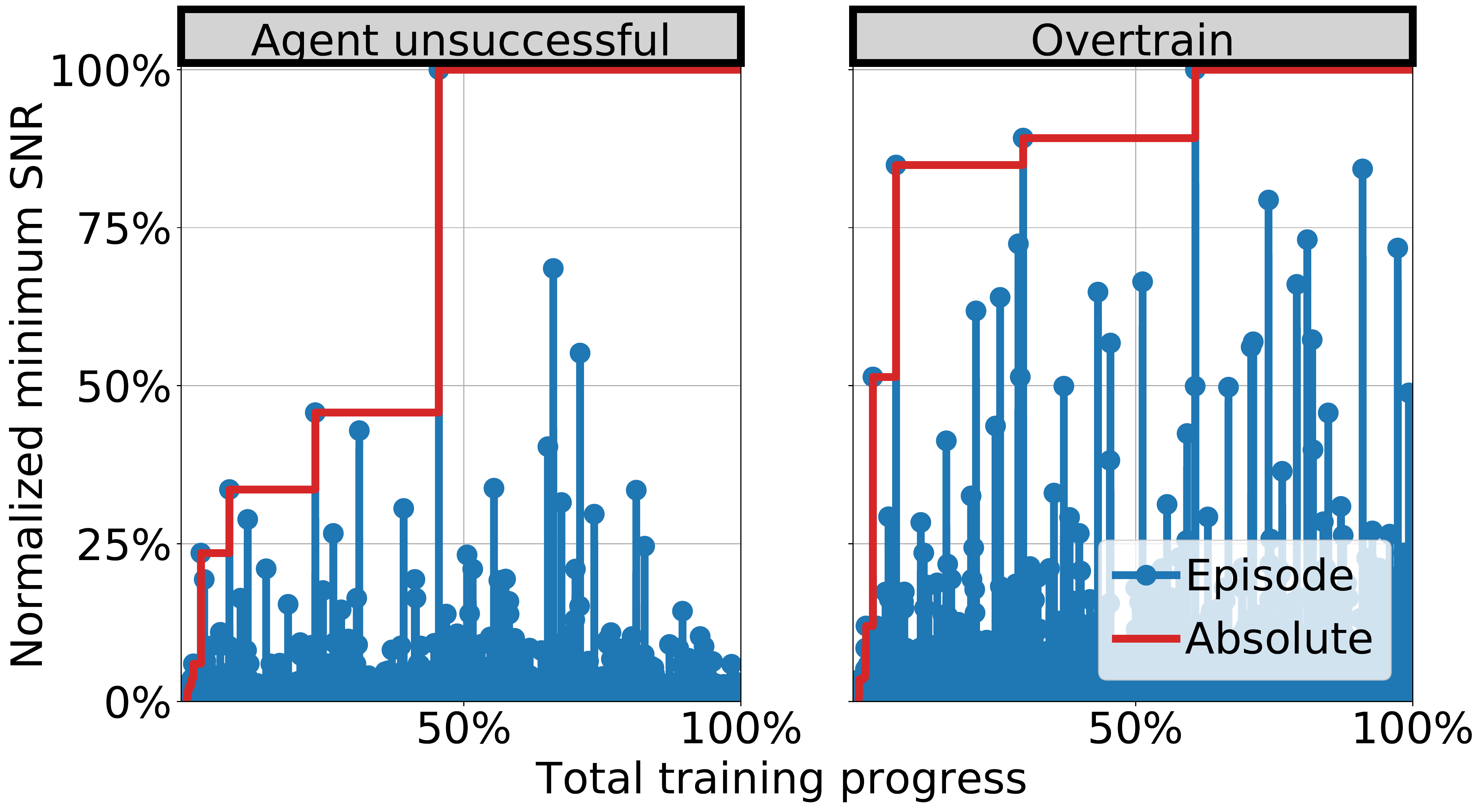}
        \caption{{Solution evolution in terms of normalized minimum \gls{snr} for a different number of deployed \glspl{ris}: (left) unsuccessful \name{} agent and (right) overtrained (right) \name{} agent.}}
        \label{fig:bad_examples} 
\end{figure}

In some instances, the agent might be unsuccessful in solving the problem at the training end just by leveraging on its own knowledge (exploitation){, as depicted in the left side of Fig.~\ref{fig:bad_examples}}. 
Nonetheless, the overall training process is still able to produce a good solution for the \gls{ris} deployment problem by combining the exploration and exploitation strategies. Once the training solution convergence is achieved, additional training time would yield a relative minimal improvement w.r.t. the solution: the fully trained agent does not further improve its output by adding more training effort, as shown on the right side of Fig.~\ref{fig:bad_examples}. 
This implies that the most value of \name{} is obtained early in the training, thus \emph{allowing for shorter training times and reducing the computational complexity}.

\section{Related work}

The process of deploying cellular networks has been extensively investigated ~\cite{Amaldi2003,albanese2022loko,Andrews2011}, but \gls{ris}-enabled network deployment is still an open problem and only a few solutions have been proposed. The optimal \gls{ris} deployment constitutes a difficult, non-convex problem due to lack of a-priori knowledge on \gls{csi} at the planning stage, which generates a \textit{deadlock}: to obtain \gls{ris} configurations we need estimations of the \gls{bs}-\gls{ris}-\gls{ue} channels whereas prior information on \gls{ris} configurations is needed to derive the reflection beampattern and so their coverage areas. Even further, optimal \gls{ris} configurations might not be able to compensate suboptimal \gls{ris} placements: breaking \gls{los} propagation conditions between \glspl{ris} and \glspl{bs} or \glspl{ue}, or resulting in deployments where \glspl{ris} serve too many \glspl{ue} that are spread on a wide area, thus reducing the achievable gain~\cite{albanese2021rennes}. In this context, simplifying assumptions on the geometric \gls{csi} can be made, reducing the problem to a more conventional deployment formulation, which can be tackled with techniques akin to well-known \glspl{bs} deployment approaches~\cite{Moro2021}. 

\gls{ai} techniques are able to deliver good-quality solutions for highly complex problems and have been proven effective in the context of \glspl{ris} and, more broadly, \gls{mimo} systems, to derive optimal beampatterns~\cite{xiong2019deep,kaur2021machine,lin2021fueling,Yang2021,zhang2021reinforcement}. Likewise, several works in the literature employ \gls{drl} approaches to solve the deployment of heterogeneous networks~\cite{Ye2020activation} or extremely dynamic scenarios such as \gls{uav}-based networks~\cite{Tarekegn2022Drones,lien2022autonomous}. 
As all of the above methods strive to produce fully-fledged \gls{drl} agents able to solve any instance of their target deployment problem, they require convergence to the optimal solution to work, which might take an arbitrarily unknown long time and extensive computing resources. This necessity can limit the applicability of the methods whenever the size of the solution (e.g., in terms of deployed devices) grows out of tractability~\cite{botvinick2019reinforcement,fenjiro2018deep}. 

\section{Conclusions}

The emerging \gls{ris} technology shows enormous potential to enable the upcoming generation of wireless networks. However, the solutions to the novel technical problems it poses, such as their intrinsic configuration, design, and deployment, need to be researched to reach a mature \gls{soa}. In particular, the deployment problem with its interdependence of the instantaneous \gls{ris} configurations unveils unprecedented challenges that were not involved while considering relays, repeaters, or the deployment of additional \glspl{bs}.

In this paper, we have proposed \name{}, a \gls{dql}-based, data-driven, model-free solution, which applies the digital twin to train and tackle the optimal deployment issue while taking into consideration the environment and the existing infrastructure. We have benchmarked \name{} against exhaustive searches and other existing solutions to the problem in a 3D model of the Rennes railway station, outperforming existing solutions by about $10$ dB in terms of minimum \gls{snr}.

\section{Acknowledgements}

This work was supported by EU H2020 RISE-6G (grant agreement 101017011) and EU H2020 METAWIRELESS (grant agreement 956256) projects. 

\bibliographystyle{IEEEtran}
\bibliography{IEEEabrv,bibliography}

\end{document}